\documentclass[preprint,prd,aps,preprintnumbers,superscriptaddress,bibnotes,nofootinbib]{revtex4-1}

\usepackage{hyperref}
\usepackage{graphicx}
\usepackage{svg}
\usepackage{float}
\usepackage{tabularx}
\usepackage{supertabular}
\usepackage{tablefootnote}
\usepackage{fancyhdr}
\usepackage{mathtools}
\usepackage{enumerate}
\usepackage{amsmath}
\usepackage{hhline}
\usepackage{amssymb,amsfonts}
\usepackage{amsthm}
\newcommand{\tsp}{\vphantom{i_1(n+\hat{1})}}

\usepackage{dcolumn}
\usepackage{bm}
\newcounter{subfigure}[figure] 
\renewcommand{\thesubfigure}{(\alph{subfigure})}
\makeatletter
\let\@makefntextOrig\@makefntext
\def\@makefntext#1{\@makefntextOrig{\baselineskip=13pt #1}}
\makeatother
\begin{document}
\title{\boldmath Triad representation for the anisotropic tensor renormalization group in four dimensions}

\author{Yuto Sugimoto\:}
\email[E-mail: ]{sugimoto@nucl.phys.tohoku.ac.jp}
\affiliation{Department of Physics, Tohoku University, Sendai 980-8578, Japan}
\author{Shoichi~Sasaki\:}
\email[E-mail: ]{ssasaki@nucl.phys.tohoku.ac.jp}
\affiliation{Department of Physics, Tohoku University, Sendai 980-8578, Japan}

\date{\today}
\begin{abstract}
The development of tensor renormalization group (TRG) algorithm in higher dimensions is an important and urgent task, as the TRG is expected to provide a way to overcome the sign problem in lattice quantum chromodynamics (QCD) calculations at finite density.
One possible approach that enables faster computations in four-dimensional lattice theories is the anisotropic tensor renormalization group (ATRG).
However, the computational cost remains substantial and requires significant computational resources.
In this paper, we propose a novel algorithm, called the triad-ATRG, which is based on the ATRG and other improved TRG variants with triad network representation.
This method achieves lower scaling with respect to the bond dimension, while minimizing the loss of accuracy in the free energy and other physical quantities.
We also present parallel implementations of both the ATRG and triad-ATRG on multiple GPUs, which significantly improve performance compared to CPU-based calculations for the four-dimensional system.

\end{abstract}
\maketitle
\section{Introduction}
\label{sec:intro}
The tensor renormalization group (TRG)~\cite{levinTensorRenormalizationGroup2007} is a powerful numerical method for computing the partition function of quantum many-body systems.
The core idea behind the TRG is to apply a sequence of low-rank approximations using singular value decomposition (SVD).
The accuracy of the approximation is controlled by the bond dimension $D$, which determines the number of singular values retained.
This procedure reduces computational complexity while preserving the essential physical information of the system.
One of the most important advantages of the TRG framework is that its computational cost is proportional to $O(\log{V})$ with respect to the system volume $V$, making it easy to take the thermodynamic limit. 
Furthermore, the absence of the sign problem is expected to achieve applications in a wide range of theories, including finite-density QCD. At present, many applications of the TRG method and its variants have been reported~
\cite{Shimizu:2012wfa,Shimizu:2012zza,Yu:2013sbi,shimizuGrassmannTensorRenormalization2014,Shimizu:2014fsa,Unmuth-Yockey:2014afa,Zou:2014rha,wangPhaseTransitionsFerromagnetic2014,takedaGrassmannTensorRenormalization2015,Yang:2015rra,nakamotoComputationCorrelationFunctions2016,zhaoTensorNetworkAlgorithm2016,kawauchiTensorRenormalizationGroup2016,Bal:2017mht,Sakai:2017jwp,sakaiApplicationTensorNetwork2018,Kadoh:2018hqq,Shimizu:2017onf,Kadoh:2018tis,Kuramashi:2018mmi,yoshimuraCalculationFermionicGreen2018,Akiyama:2019chk,Bazavov:2019qih,akiyamaPhaseTransitionFourdimensional2019,Kadoh:2019ube,akiyamaTensorRenormalizationGroup2020,Delcamp:2020hzo,Iino:2019rxt,PhysRevE.101.012124,akiyamaTensorRenormalizationGroup2021,Akiyama:2020soe,Akiyama:2021zhf,Bloch:2021mjw,Fukuma:2021cni,hirasawaTensorRenormalizationGroup2021a,Milde:2021vln,Nakayama:2021iyp,kuwaharaTensorRenormalizationGroup2022,Akiyama:2021glo,Akiyama:2022pse,Akiyama:2022eip,Jha:2022pgy,Luo:2022eje,Akiyama:2023hvt,Akiyama:2024qgv,Samlodia:2024kyi,Akiyama:2024qer,Pai:2024tip,kannoGrassmannTensorRenormalization2024,luoCriticalEndpointsThreedimensional2025a,tsengBondWeightedTensor2025}.

From an algorithmic perspective, the original TRG was designed for two-dimensional classical systems, and its extension to higher dimensions is not straightforward.
The higher-order tensor renormalization group (HOTRG)~\cite{xieCoarsegrainingRenormalizationHigherorder2012} was developed to extend the TRG to higher dimensions by implementing coarse-graining transformations along a selected axis at each step.
Recently, the HOTRG has been successfully applied to four-dimensional systems~\cite{akiyamaPhaseTransitionFourdimensional2019,Akiyama:2019chk,Milde:2021vln}. These achievements are of particular importance, as our target theory, QCD, is defined in $(3\!+\!1)$-dimensional spacetime.
However, due to its relatively simple algorithm structure, the HOTRG suffers from high computational cost in higher dimensions, since its computational cost is proportional to $O(D^{4d-1})$ in $d$ dimensions.
As a result, increasing the bond dimension in three and four-dimensional systems poses a significant computational challenge.

In order to address this situation, several variants of the HOTRG have been proposed to reduce the computational cost in higher dimensions. Among these methods, the anisotropic TRG (ATRG)~\cite{adachiAnisotropicTensorRenormalization2020} has been demonstrated to achieve a substantial cost reduction, scaling as $O(D^{2d+1})$. This lower cost enables the study of four-dimensional systems with substantially larger bond dimensions, and in fact, the ATRG has been successfully applied to four-dimensional systems with/without the sign problem~\cite{Akiyama:2019chk,akiyamaTensorRenormalizationGroup2020,Akiyama:2020soe,Akiyama:2021zhf,Akiyama:2022eip,Akiyama:2023hvt}.

In addition to the ATRG, there exist other algorithms that have been developed for the purpose of reducing computational costs~\cite{kadohRenormalizationGroupTriad2019,Milde:2021vln,nakayamaApplicationProjectiveTruncation2024,nakayamaRandomizedHigherorderTensor2023}. It is noteworthy that the triad-TRG~\cite{kadohRenormalizationGroupTriad2019}, the triad-MDTRG~\cite{nakayamaApplicationProjectiveTruncation2024,nakayamaRandomizedHigherorderTensor2023}, achieve a lower scaling than the HOTRG by decomposing all tensors into three-legs.
The triad-TRG and a series of MDTRG variants also adopt approximate contraction schemes based on the randomized singular value decomposition (RSVD)~\cite{doi:10.1137/090771806}, which was originally introduced in the two-dimensional TRG~\cite{moritaTensorRenormalizationGroup2018}.
In particular, the MDTRG incorporates several strategies so as to suppress additional systematic errors introduced by the tensor decomposition. As a result, the MDTRG achieves an accuracy comparable to that of the HOTRG when simulating the three-dimensional Ising model. However, the application of the MDTRG to four-dimensional systems remains unreported.

In this study, we propose a novel approach, the triad-ATRG, which combines the ATRG and the triad representation~\cite{kadohRenormalizationGroupTriad2019,nakayamaRandomizedHigherorderTensor2023} to enable practical computations with larger bond dimensions for four-dimensional systems.
We also present parallel implementations of both the ATRG and triad-ATRG on multiple GPUs, which significantly improve performance compared to CPU-based calculations.

This paper is organized as follows.
In Sec.~\ref{sec:triad_representation}, we review the basics of both the HOTRG and ATRG methods, as well as the triad representation. Then, we introduce the triad-ATRG method, which is more efficient than the original ATRG method
for the case of the four-dimensional system.
The idea of combining the triad-ATRG and the MDTRG technique to achieve a higher accuracy calculation is also described.
We also discuss the GPU parallelization for the ATRG and triad-ATRG methods.
In Sec.~\ref{sec:numerical_results}, we present the numerical results of applying the triad-ATRG method to the four-dimensional Ising model. We then demonstrate that the triad-ATRG method achieves a comparable accuracy to the ATRG method, but with a significantly reduced computational cost.
In Sec.~\ref{sec:conclusion}, we summarize our work and discuss future prospects.

\section{triad-ATRG method}
\label{sec:triad_representation}
\subsection{Overview of the ATRG, MDTRG, and triad representation}

In this section, we briefly review the anisotropic tensor renormalization group (ATRG) method~\cite{adachiAnisotropicTensorRenormalization2020}.
The ATRG method is based on the higher-order tensor renormalization group (HOTRG) method~\cite{xieCoarsegrainingRenormalizationHigherorder2012}, with some additional approximation techniques.

Let us consider a target tensor network that is connected on a $d$-dimensional square lattice on the volume $V=L^d$.
We assume that this tensor network is homogeneous, implying that all tensors located at each lattice site are equivalent.
This condition is satisfied when we consider a partition function of a quantum many-body system, which has a local translational invariance. In such case, the {partition} function of the system is expressed as
%
%
\begin{equation}\label{eq:Z}
    Z = \mathrm{tTr} \prod_n T_{i_1(n)i_2(n)\dots i_d(n)j_1(n)j_2(n)\cdots j_d(n)}.
\end{equation}
where $n$ denotes a lattice site. In the above equation, the symbol $T$ represents a local tensor; $i_\mu(n)$ corresponds to the bond connecting site $n$ to $n+\hat{\mu}$, while $j_\mu(n)$ is a shorthand for the opposite bond connecting site $n$ to $n-\hat{\mu}$, \emph{i.e.}, $j_\mu(n) \equiv i_\mu(n-\hat{\mu})$.
The tensor trace $\mathrm{tTr}$ indicates the summation over all internal indices under the constraint $i_\mu(n) = j_\mu(n+L\hat{\mu})$ for $\mu = 1, 2, \dots, d$ under periodic boundary conditions. 

\begin{figure}[t]
    \centering
    \includegraphics[width=0.85\textwidth]{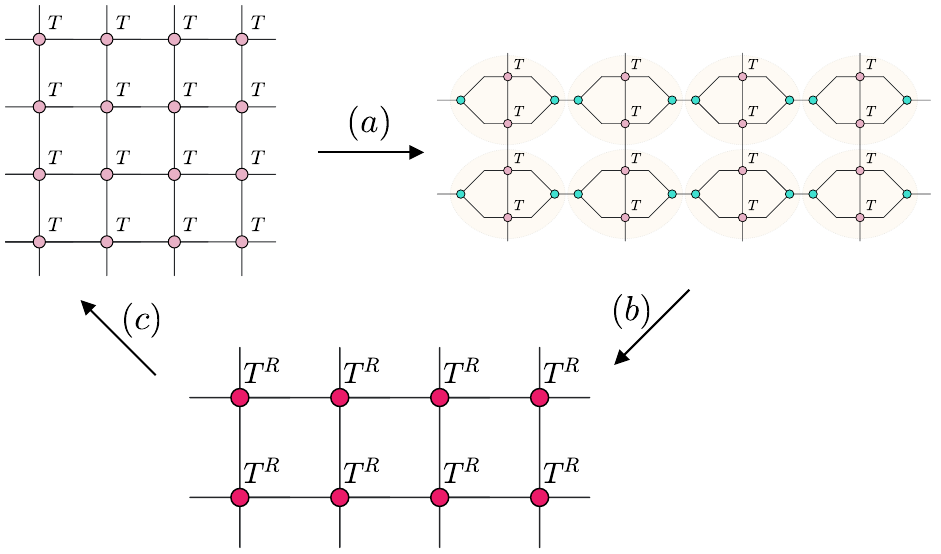}
    \caption{
        Illustration of the HOTRG method on a two-dimensional square lattice. {The procedure is as follows:}
        (a) Apply a low-rank projector constructed from the HOSVD of $\Gamma$.
        (b) Perform tensor contractions around the yellow patches.
        (c) Rotate the system and repeat the procedure.
    }\label{fig::HOTRG}    
\end{figure}

A fundamental technique common to all TRG methods is {the approximation of} the target tensor network by updating the local tensors through SVD.
In the HOTRG case, the approximation is performed along a selected axis by contracting the unit-cell tensor, denoted by $\Gamma$, {which is} composed of two adjacent local tensors.
The unit-cell tensor along $\hat{1}$-axis is defined by the contraction {with respect to $i_1(n)=j_1(n+\hat{1})$} as follows:
%
%
\begin{equation}
    \Gamma=\sum_{i_1(n)} T_{i_1(n+\hat{1})i_2(n+\hat{1})\dots i_d(n+\hat{1})i_1(n)j_2(n+\hat{1})\dots j_d(n+\hat{1})}T_{i_1(n)i_2(n)\dots i_d(n)j_1(n)j_2(n)\dots j_d(n)}.
\end{equation}
In the HOTRG method, the approximation based on the higher-order singular value decomposition (HOSVD)~\cite{delathauwerMultilinearSingularValue2000} 
{is applied to} each axis of $\Gamma$.
In this approximation, The $D$-largest singular values are maintained, while the remaining small singular values are truncated.
The truncated SVD yields the projectors $U^{(\alpha)}\,(\alpha=2,3,\dots,d)$, which span the dominant subspaces of $\Gamma$.
A low-rank approximation of $\Gamma$ is accomplished by inserting $U^\dagger U \approx I$ at each relevant bond. 
This procedure allows us to define a new renormalized tensor, $T^R$, by contracting the unit-cell tensor $\Gamma$ with a surrounding set of projectors. Specifically, the coarse-graining process along the $\hat{1}$-direction is expressed as:
%
%
\begin{align}\label{eq:HOTRG}
    &T_{i_1(n+\hat{1})k_2\dots k_d j_1(n)k_2'\dots k_d'}^{R} 
    \nonumber \\
    &= \hspace{-10pt}\sum_{/ i_1(n+\hat{1}), j_1(n),\{k\},\{k^\prime\}}  \hspace{-10pt}T_{i_1(n+\hat{1})i_2(n+\hat{1})\dots i_d(n+\hat{1})i_1(n)j_2(n+\hat{1})\dots j_d(n+\hat{1})}T_{i_1(n)i_2(n)\dots i_d(n)j_1(n)j_2(n)\dots j_d(n)}
    \nonumber\\
    &\hspace{50pt} \quad \times U_{i_2(n+\hat{1}) i_2(n)k_2}^{*(2)} U_{j_2(n+\hat{1}) j_2(n)k_2'}^{(2)}\cdots U_{i_d(n+\hat{1})i_d(n)k_d}^{*(d)} U_{j_d(n+\hat{1})j_d(n)k_d'}^{(d)} 
\end{align}
where $T^R$ denotes the renormalized tensor, and indices $k_\mu$ and $k'_\mu$ represent the new bonds created {by this contraction. Here, the notation $/ i_1(n+\hat{1}),{j_1(n)},\{k\},\{k^\prime\}$ denotes a summation that is taken over all indices except for $i_1(n+\hat{1}),{j_1(n)},k_2,\dots ,k_d,k^\prime_2,\dots ,k^\prime_d$.} 
The sequence of procedures by which the unit cell tensor, originally composed of two adjacent fundamental tensors $T$, is contracted into a single renormalized tensor $T^R$ is referred to as {\it the coarse-graining step}.
As a result, the number of tensors along the $\hat{1}$-axis {is reduced by half, resulting in the creation of} a new coarse lattice.
After repeating this process for each axis, an approximate renormalized tensor network can be obtained.
Figure~\ref{fig::HOTRG} illustrates the algorithm flow of the HOTRG method.
The computational cost of the procedure described in Eq.~\eqref{eq:HOTRG} is $O(D^{4d-1})$. This becomes the bottleneck in the application of the HOTRG method to higher-dimensional systems.
Due to this high computational cost, increasing the bond dimension in four or higher dimensions poses a substantial challenge.

The ATRG method reduces the computational cost to $O(D^{2d+1})$ by applying the low-rank approximation for $\Gamma$, as described below. We follow the modified algorithm proposed in Ref.~\cite{oai:tsukuba.repo.nii.ac.jp:02005504}.
Let us consider the coarse-graining process along the $\hat{1}$-direction.
The process consists of two additional steps. 
At first, $\Gamma$ is approximated by $\Gamma^{\text{ATRG}}$ which is the contraction of four low-rank tensors $A,B,C$, and $D$ with the truncated bond dimension $D$ as follows:
%
%
\begin{align} 
\label{eq:ABCD}
\Gamma &\approx\Gamma^{\text{ABCD}}\nonumber\\
&\coloneqq\sum_{\alpha,\beta,i_1(n)}^{D}A_{i_1(n+\hat{1})i_2(n+\hat{1})\dots i_d(n+\hat{1})\alpha}B_{i_1(n)j_2(n+\hat{1})\dots j_d(n+\hat{1})\alpha}C_{i_1(n)i_2(n)\dots i_d(n)\vphantom{i_1(n+\hat{1})}\beta}D_{j_1(n)j_2(n)\dots j_d(n)\vphantom{i_1(n+\hat{1})}\beta},
\end{align}
where the bond dimension and four tensors are
introduced by the truncated SVD of the tensor $T$\footnote{
It is generally recommended to derive the tensors $A$, $B$, $C$, and $D$ from the SVD of $\Gamma$ in order to reduce the systematic error associated with the truncation of the partition function $Z$.
In the present study, however, we employ the SVD of $T$. 
This is because the decomposition of Eq.~\eqref{eq:ABCD} can be performed exactly, without truncation, for Ising model calculations with the truncated bond dimension $D$ greater than $2^d$.
}. 

The second step of the ATRG, also known as {\it the bond-swapping step}, involves an additional approximation that is applied prior to the coarse-graining step.
The bond-swapping step is defined as the process of swapping the forward bond indices, $i_2(n),\dots ,i_d(n)$ of $C$
and the backward bond indices, $j_2(n+\hat{1}),\dots ,j_d(n+\hat{1})$ of $B$ by employing the following truncated SVD:
%
%
\begin{align}
\sum_{i_1(n)}^D B_{i_1(n)j_2(n+\hat{1})\dots j_d(n+\hat{1})\alpha}C_{i_1(n)i_2(n)\dots i_d(n)\vphantom{i_1(n+\hat{1})}\beta}\approx 
\sum_{\gamma}^{D}X_{\alpha i_2(n)\dots i_d(n)\gamma}\sigma_{\gamma\gamma}Y_{\beta j_2(n+\hat{1})\dots j_d(n+\hat{1})\gamma},
\end{align}
where $X$ and $Y$ are isometric tensors and $\sigma$ is the diagonal matrix of singular values, which
are truncated, except for the $D$-largest values. Using this decomposition, the unit-cell tensor $\Gamma$
is approximated by $\Gamma^{\text{ATRG}}$ defined as 
\begin{figure}[t]
	\centering
	\includegraphics[width=0.95\textwidth]{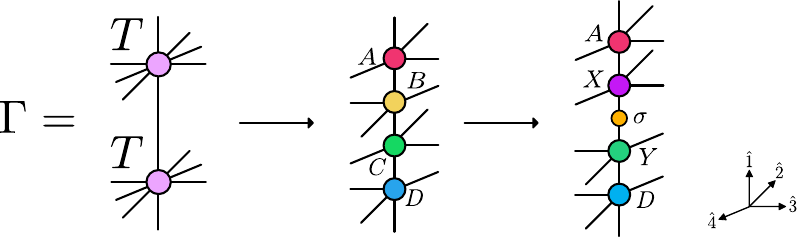}
	\caption{The decomposition of the tensor $\Gamma$ in the ATRG. The left figure shows $\Gamma$ in four dimension, and the center figure illustrates the Eq.~\eqref{eq:ABCD}, and right figure illustrates the Eq.~\eqref{eq:GammaATRG}}\label{fig::gammaATRG}
\end{figure}
%
%
\begin{align}\label{eq:GammaATRG}
    \Gamma^{\text{ABCD}}&\approx\Gamma^{\text{ATRG}}\nonumber\\
    &\coloneqq\sum_{\alpha,\gamma,\beta}^D A_{i_1(n+\hat{1})i_2(n+\hat{1})\dots i_d(n+\hat{1})\alpha}X_{\alpha i_2(n)\dots i_d(n)\vphantom{i_1(n+\hat{1})}\gamma} \nonumber\\
&\hspace{60pt}\times \sigma_{\tsp\gamma\gamma} Y_{\beta j_2(n+\hat{1})\dots j_d(n+\hat{1})\gamma}D_{j_1(n)j_2(n)\dots j_d(n)\vphantom{i_1(n+\hat{1})}\beta}.
\end{align}
Figure~\ref{fig::gammaATRG} schematically illustrates how the tensor $\Gamma$ is factorized in four dimensions.

As detailed in Appendix~\ref{append:A}, the computational cost of the bond-swapping step is $O(q\eta D^6)$, where $q$ and $\eta$ represent the number of QR decompositions and the oversampling parameter in the randomized SVD method~\cite{10.1093/ptep/ptz133}, respectively. The coarse-graining process of the ATRG method is accomplished by applying squeezers $E^{(\alpha)},F^{(\alpha)}\,(\alpha=2,3,\dots , d)$ to the tensor $\Gamma$,
%
%
\begin{align}
	\label{makeG} 
	\hspace{-18pt}G_{i_1k_2\cdots k_d\gamma}&=\sum_{/\gamma,i_1,\{k\}}A_{i_1i_2(n+\hat{1})\dots i_d(n+\hat{1})\alpha}X_{\alpha i_2(n)\dots i_d(n)\vphantom{i_1(n+\hat{1})}\gamma}
    \sigma_{\tsp\gamma\gamma}
E^{(2)}_{\smash[t]{i_2(n+\hat{1}) i_2(n)k_2}}\dots E^{(d)}_{\smash[t]{i_d(n+\hat{1}) i_d(n)k_d}}\\
	\label{makeH} \hspace{-18pt}H_{\tsp j_1k_2'\cdots k_d' \gamma}&=\sum_{/\gamma,j_1,\{k'\}}D_{j_1j_2(n)\dots j_d(n)\vphantom{i_1(n+\hat{1})}\beta}Y_{\beta j_2(n+\hat{1})\dots j_d(n+\hat{1})\gamma}F^{(2)}_{\smash[t]{j_2(n) j_2(n+\hat{1})k_2'}}\dots F^{(d)}_{\smash[t]{j_d(n) j_d(n+\hat{1})k_d'}},
\end{align}
where the notation $/\gamma,i_1,\{k\}$ denotes a summation that is taken over all indices except for $\gamma,i_1,k_2,\dots ,k_d$.
These squeezers are derived in each direction, thereby achieving the lower bound of the Frobenius norm with the adjacent unit cell $\Gamma$~\cite{wangClusterUpdateTensor2011,PhysRevB.90.174201,iinoBoundaryTensorRenormalization2019,PhysRevD.110.094501}.
Due to several approximations, the ATRG significantly reduces the computational cost of the coarse-graining process to $O(D^{2d+1})$.  
This reduction is achieved through the low-rank representation of the local tensor $\Gamma$, in which bond swapping is employed to split the tensor contraction into two independent parts, as given in Eqs.~\eqref{makeG} and~\eqref{makeH}.  
As a result, the number of isometries (squeezers) required in each contraction is reduced by half compared to the HOTRG method.

An alternative representation of $\Gamma$, known as the triad representation, was first proposed in Ref.~\cite{kadohRenormalizationGroupTriad2019} and later extended in Ref.~\cite{nakayamaRandomizedHigherorderTensor2023}.
The central idea of the triad representation is to decompose all tensors into contractions of three-leg tensors.
In the triad-MDTRG method, pairs of isometries are constructed via the truncated SVD of $\Gamma$ which  reduces systematic error for the partition function $Z$ because $\Gamma$ is a larger network than $T$.\footnote{In the triad-TRG method~\cite{kadohRenormalizationGroupTriad2019}, the triad representation is constructed from the decomposition of the tensor $T$.
As pointed out later in Ref.~\cite{nakayamaRandomizedHigherorderTensor2023}, 
the application of the decomposition of $\Gamma$ as employed in the HOTRG method, results in a faster decay of the singular value spectrum and yields more accurate results.
On the other hand, the bond-swapping step in the ATRG can, in principle, be interpreted as a decomposition of the unit-cell tensor $\Gamma$, since the constituent tensors $A$ and $D$ are isometric. 
Although the original formulation of the ATRG~\cite{adachiAnisotropicTensorRenormalization2020} did not exploit such a broader decomposition to construct the squeezers, it is straightforward to consider such decompositions taking into account the canonical structure of $\Gamma^{\text{ATRG}}$ (see Sec.~\ref{sec:triad-ATRG}). Such modification was proposed in Ref.~\cite{akiyamaTensorRenormalizationGroup2020}. In the ATRG implementation used in this study, the SVD of $\Gamma$ is adopted following Ref.~\cite{oai:tsukuba.repo.nii.ac.jp:02005504} in all steps except for the construction of the initial tensors $A$, $B$, $C$, and $D$.
}
In this method, when the bond dimension is set to $D$, it is enlarged to $rD$ for internal lines, where $r$ is an integer parameter that we refer to as the \textit{overspanning parameter}\footnote{
The parameter $r$ originates from the oversampling parameter used in the RSVD part of the MDTRG method.
Although it was originally called the oversampling parameter, we prefer to use the term ``overspanning parameter'' here to avoid confusion. This emphasizes that the $rD$ singular values are retained in order to \textit{enlarge the spanned subspace}.
}.

\begin{figure}[htbp]
\centering
\includegraphics[width=0.55\textwidth]{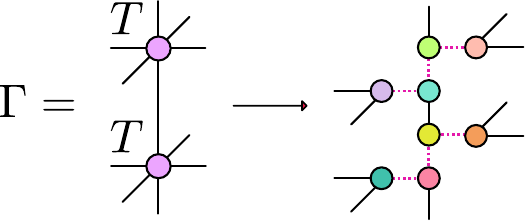}
\caption{Schematic illustration of the triad representation of $\Gamma$ in the three-dimensional case.  
The unit-cell tensor $\Gamma$ (left) is expressed in the triad representation (right).  
The bond dimension of the dotted lines is overspanned to $rD$, while that of the solid lines is set to $D$.}\label{fig::MDTRG}
\end{figure}

On the other hand, the MDTRG and its variants also employ approximation via RSVD in the coarse-graining process, in contrast to the ATRG.
As a result, in the tirad representation, the computational cost is reduced to $O(D^{d+3})$ multiplied by an $r$-dependent factor arising from the triad representation and the internal-line overspanning, while maintaining accuracy comparable to that of HOTRG.
The schematic figure of the triad representation in the triad-MDTRG is illustrated for a three-dimensional case in Fig.~\ref{fig::MDTRG}. 
The prescription of the triad representation can also be applied to any type of tensor network, including the ATRG method.
\subsection{triad-ATRG}\label{sec:triad-ATRG}
In this subsection, we explain how to reduce the computational cost of the ATRG by employing its triad representation. In this paper, we particularly focus on the four-dimensional case. Note that this algorithm is based on our previous proceedings~\cite{Sugimoto:2025jK}.
First, we demonstrate how to construct the triad representation on the unit-cell tensor after the bond-swapping step in the ATRG method, (denoted by $\Gamma^{\text{ATRG}}$), and then outline how the algorithm is modified.

According to the MDTRG method~\cite{nakayamaRandomizedHigherorderTensor2023}, pairs of isometries are constructed by performing a truncated HOSVD on $\Gamma^{\text{ATRG}}$.
For example, the isometric tensor $U^{A}$ associated with the tensor $A$ is obtained as follows:
%
%
\begin{align}\label{eq:HOSVD}
    \hspace{-10pt}(AX\sigma A^\dagger X^\dagger\sigma^\dagger)_{i_2(n+\hat{1}) i_3(n+\hat{1}) \overline{i_2}(n+\hat{1}) \overline{i_3}(n+\hat{1})}\approx\sum_{k_A}^{rD}U_{i_2(n+\hat{1}) i_3(n+\hat{1})k_A}^{A}\left(S_{\tsp k_Ak_A}^{A}\right)^2 U_{{\overline{i_2}(n+\hat{1})\overline{i_3}(n+\hat{1})k_A}}^{*A},
\end{align}
where the internal lines of these isometries are overspanned as in the case of the MDTRG. Here, the bonds of the conjugate tensor $A^\dagger$ are represented as $\overline{i_\mu}(n+\hat{1}),\overline{i_\mu}(n)$ for $\mu=1,\,2,\,3,\,4$.
As described in Eq.~\eqref{eq:ABCD}, the unit-cell tensor after the bond-swapping step in the ATRG method is expressed as $\Gamma^{\text{ATRG}} = A X \sigma Y \!D$, where the tensors $A$, $X$, $Y$, and $D$ are all isometric.  
Since each of these tensors acts as an identity when contracted with its conjugate during the SVD process, only $A$, $X$, and $\sigma$ are required to construct $U^A$.
This structure is known as the canonical form of the ATRG~\cite{Akiyama:2024qgv}, and it simplifies the practical computation.  
The other isometries, $U^D$, $U^X$, and $U^Y$, can be derived in the same manner.

\begin{figure}[t]
	\centering
	\includegraphics[width=0.85\textwidth]{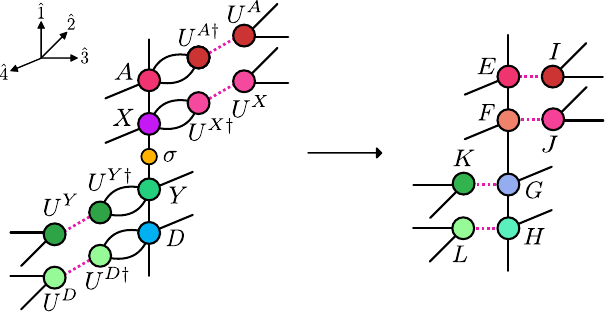}
	\caption{The triad representation of $\Gamma$. The left figure shows $\Gamma$ in four dimension, and the right figure illustrates the triad representation. Note that the bond dimension of magenta dotted lines are overspanned from $D$ to $rD$. }\label{fig::triadrep}
\end{figure}
After multiplying by each isometry to the respective tensor, we get the triad representation of $\Gamma^{\text{ATRG}}$ expressed by the following new tensors,
%
%
\begin{align}
    &E_{i_1(n+\hat{1})k_A i_4(n+\hat{1})\alpha}=\sum_{i_2(n+\hat{1}),i_3(n+\hat{1})}A_{i_1(n+\hat{1})i_2(n+\hat{1})i_3(n+\hat{1}) i_4(n+\hat{1})\alpha}U_{\smash[t]{i_2(n+\hat{1}) i_3(n+\hat{1})k_A}}^{*A} \label{eq:all_fundamental_tensors_E}\\
    &F_{\alpha k_X i_4(n)\tsp\gamma}=\sum_{i_2(n),i_3(n)}X_{\alpha i_2(n)i_3(n) i_4(n)\vphantom{i_1(n+\hat{1})}\gamma}U_{\smash[t]{i_2(n) i_3(n)k_X}}^{*X}\sigma_{\gamma\gamma}\\
    &G_{\beta k_Y j_4(n+\hat{1})\gamma}=\sum_{j_2(n+\hat{1}),j_3(n+\hat{1})}Y_{\beta j_2(n+\hat{1})j_3(n+\hat{1}) j_4(n+\hat{1})\gamma}U_{\smash[t]{j_2(n+\hat{1}) j_3(n+\hat{1})k_Y}}^{*Y}\\
    &H_{j_1(n)k_D j_4(n)\tsp\beta}=\sum_{j_2(n),j_3(n)}D_{j_1(n)j_2(n)j_3(n) j_4(n)\vphantom{i_1(n+\hat{1})}\beta}U_{\smash[t]{j_2(n)j_3(n)k_D}}^{*D}\\
    &I_{i_2(n+\hat{1}) i_3(n+\hat{1})k_A}=U_{\smash[t]{i_2(n+\hat{1}) i_3(n+\hat{1})k_A}}^{A}\\
    &J_{\tsp i_2(n) i_3(n)k_X}=U_{\smash[t]{i_2(n) i_3(n)k_X}}^{X}\\
    &K_{j_2(n+\hat{1}) j_3(n+\hat{1})k_Y}=U_{\smash[t]{j_2(n+\hat{1}) j_3(n+\hat{1})k_Y}}^{Y}\\
    &L_{\tsp j_2(n)j_3(n)k_D}=U_{\smash[t]{j_2(n)j_3(n)k_D}}^{D}.
    \label{eq:all_fundamental_tensors_L}
\end{align}
These tensors are schematically illustrated in Fig.~\ref{fig::triadrep}.  
The new unit-cell tensor, $\Gamma^{\rm Triad}$, is represented by four-leg tensors $E$, $F$, $G$, and $H$, and three-leg tensors, $I$, $J$, $K$, and $L$.
This step of introducing the triad representation into the ATRG method requires a cost of $O(D^7)$ to perform the contraction of Eq.~\eqref{eq:HOSVD} in the four-dimensional case.

In the next step, we introduce the squeezers $M^{(\alpha)},N^{(\alpha)}\,(\alpha=2,3,\dots, d)$ that are derived from $\Gamma^{\text{Triad}}$. 
In the coarse-graining process, we apply the squeezers to $\Gamma^{\text{Triad}}$ as follows:
%
%
\begin{align}
        \label{makeMG} \MoveEqLeft[35]
        \Phi_{i_1k_2k_3 k_4\gamma}=\sum_{/\gamma,i_1,k}\left(E_{i_1(n+\hat{1})k_A i_4(n+\hat{1})\alpha}F_{\alpha k_X i_4(n)\tsp\gamma}M^{(4)}_{\smash[t]{i_d(n+\hat{1}) i_d(n)k_d}}\right)\nonumber\\
        \times\left(I_{i_2(n+\hat{1}) i_3(n+\hat{1})k_A}J_{\tsp i_2(n) i_3(n)k_X}M^{(2)}_{\smash[t]{i_2(n+\hat{1}) i_2(n)k_2}}M^{(3)}_{\smash[t]{i_3(n+\hat{1}) i_3(n)k_3}}\right) \\
        \label{makeMH} \MoveEqLeft[35] \Psi_{\tsp j_1k_2'k_3' k_4' \gamma}=\sum_{/\gamma,j_1,k'}\left(G_{\beta k_Y j_4(n+\hat{1})\gamma}H_{j_1(n)k_D j_4(n)\tsp\beta}N^{(4)}_{\smash[t]{j_4(n) j_4(n+\hat{1})k_4'}}\right)\nonumber\\
        \times\left(K_{j_2(n+\hat{1}) j_3(n+\hat{1})k_Y}L_{\tsp j_2(n)j_3(n)k_D}N^{(2)}_{\smash[t]{j_2(n) j_2(n+\hat{1})k_2'}}N^{(3)}_{\smash[t]{j_3(n) j_3(n+\hat{1})k_3'}}\right).
\end{align}
The introduction of the triad representation reduces the computational cost of the coarse-graining process from $O(D^9)$ to $O(r^2 D^7)$.

\begin{figure}[t]
	\centering
	\includegraphics[width=0.85\textwidth]{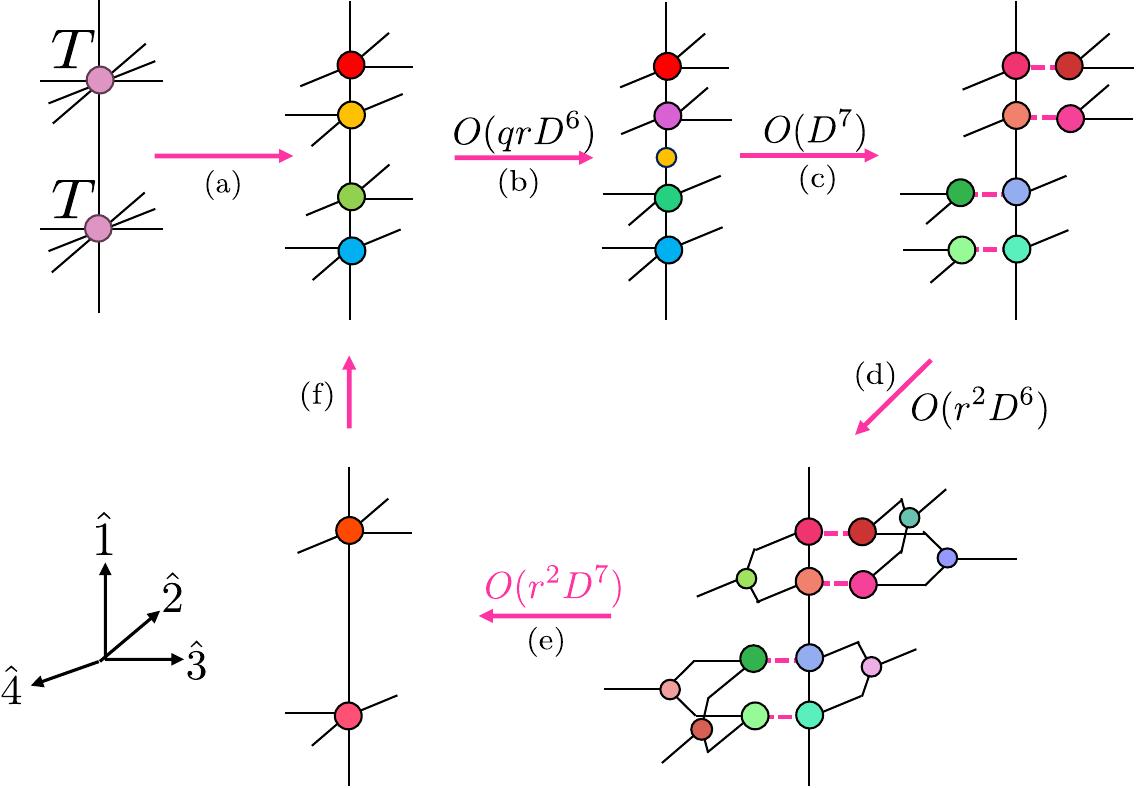}
	\caption{{Schematic overview of the triad-ATRG algorithm:} (a) Decomposition of the fundamental tensor. (b) Bond-swapping step. (c) Construction of the triad representation. (d) Derivation of the squeezer. (e) Contraction {(coarse-graining)}. (f) Moving onto the next step.}\label{fig::triadflow}
\end{figure}

Figure~\ref{fig::triadflow} finally depicts a
graphical representation of the entire triad-ATRG process.
The bottleneck of computational cost is significantly lower than that of the original ATRG method, being proportional to $O(r^2 D^7)$.~\footnote{
In our work, the oversampling parameter, $\eta$, in the RSVD is set to $\eta = O(r)$. Therefore, hereinafter, we do not distinguish between the oversampling parameter $\eta$ and the overspanning parameter $r$ in the ATRG and triad-ATRG methods. Despite the absence of the overspanning technique in the ATRG method, we set $\eta \coloneqq O(r) > 1$, which is of the same order as that used in the bond-swapping part of the triad-ATRG method.
It is also worth {noting} that according to Ref.~\cite{10.1093/ptep/ptz133}, if we set $q = O(D)$, the computational cost of the bond-swapping step becomes $O(r D^7)$, which is not negligible.
However, since this term is subleading in $r$, the contraction step, which scales as $O(r^2 D^7)$, remains the dominant contribution to the computational cost.
}
Note that when $r=D$ is chosen, the triad-ATRG method recovers the same computational scaling as the original ATRG method.

Here, we need to add a brief comment on the cases of $d \neq 4$.
For $d = 2$, the triad-ATRG is identical to the ATRG since the original ATRG is already expressed in a triad representation.
On the other hand, in the case of $d = 3$, the total computational cost in the triad-ATRG becomes $O(r^2 D^6)$, which is not significantly lower than the original ATRG cost of $O(D^7)$.  
Therefore, the practical speedup is limited in this case.  
When $d > 4$, the computational cost scales as $O(D^{d+3})$ 
with an additional factor of powers of $r$, depending on how the triad representation is adopted.
However, a complete survey of the case of $d>4$ is beyond the scope of this study. Since our primary objective is the application to QCD, the present study focuses exclusively on the implementation in four dimensions.

\subsection{Overspanning for the bond-swapping step}
The MDTRG method employs an overspanning technique for internal lines which leads to improved accuracy.  
In contrast, the conventional ATRG method does not incorporate this technique.  
Since the bond-swapping step introduces the additional systematic error from the HOTRG method, applying the overspanning technique to the bond-swapping step is expected to significantly enhance the accuracy of the triad-ATRG method.

In the original ATRG method, introducing overspanning in the bond-swapping step increases the computational cost to $\mathcal{O}(r_{\text{bond}}^2 D^9)$, where $r_{\text{bond}}$ is the overspanning parameter for the RSVD used in the bond swapping. This cost is remarkably high. Furthermore, the memory cost of the intermediate tensors $X$ and $Y$ is calculated to be $\mathcal{O}(r_{\text{bond}}^2 D^5)$~\footnote{If we introduce internal-line overspanning in the bond-swapping step at the first iteration of the ATRG, the tensors $X$ and $Y$ become of size $D \times D \times D \times D \times (r_\text{bond}D)$.  
The influence of this overspanning is carried over to the fundamental tensors $A$, $B$, $C$, and $D$ in the subsequent step, so that the subscripts $\alpha$ and $\beta$ in Eq.~\eqref{eq:ABCD} are also overspanned.  
As a result, due to this point, the tensors $X$ and $Y$ become of size $(r_\text{bond}D) \times D \times D \times D \times (r_\text{bond}D)$.}. 
This additional memory cost can also act as a significant bottleneck in practical applications.

In this paper, we incorporate internal-line overspanning (ILO) into the triad-ATRG method for the bond-swapping part as well as the triad representation. This method, referred to as the ILO triad-ATRG method, offers higher accuracy and lower computational costs.

Two major computational bottlenecks in the ILO triad-ATRG method are the bond-swapping and coarse-graining steps, whose costs are $\mathcal{O}(qr_{\text{bond}}^4 D^6 + r^2 r_{\text{bond}}^2 D^7)$ and remain practically feasible.  
The coarse-graining procedure is schematically illustrated in Fig.~\ref{fig::ILOtriad}, while the details of the bond-swapping process are provided in the Appendix \ref{append:A}.

In addition to the reasonable computational cost described above, QR decomposition is applied before the bond-swapping step in practical implementations~\cite{oai:tsukuba.repo.nii.ac.jp:02005504},  
which allows us to avoid constructing large intermediate tensors $X$ and $Y$, and to directly obtain the triad representation of $\Gamma^{\text{ATRG}}$.~\footnote{
In the original ATRG method, the construction of large intermediate tensors $X$ and $Y$ can also be avoided by carefully employing the memory blocking technique~\cite{moritaTensorRenormalizationGroup2018} in both the squeezer step and the coarse-graining step.  
However, the computational cost remains substantial even with this technique.}
\enlargethispage{\baselineskip}

\begin{figure}[htbp]
\centering
\includegraphics[width=0.65\textwidth]{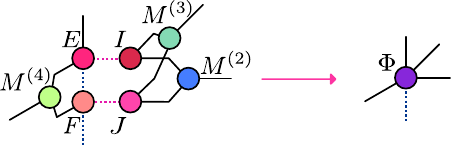}
\caption{Schematic view of the coarse-graining step in the ILO triad-ATRG procedure. Magenta and blue dotted lines are overspanned to $rD$ and $r_{\text{bond}}D$, respectively. This procedure requires an $O(r^2 r_{\text{bond}}^2 D^7)$ cost.}\label{fig::ILOtriad}
\end{figure}

\subsection{GPU parallelization}
The TRG framework primarily consists of tensor contraction and SVD, both of which are based on matrix multiplication.
Therefore, the TRG is suitable for GPU acceleration because both operations can be easily parallelized.
This results in significant speedups compared to CPU-based implementations. In the case of the HOTRG method, GPU parallelization has already been successfully applied to two-dimensional systems~\cite{jhaGPUAccelerationTensorRenormalization2024}. However, due to the substantial increase in memory requirements and availability, it remains challenging to extend this approach to three- and four-dimensional cases.

In this work, we have successfully applied GPU parallelization to the implementation of both the ATRG and triad-ATRG methods by distributing the contraction and SVD operations across multiple GPUs.
In the following, we assume that $n$ GPUs are available and
denote $i$-th device as GPU $i$.

The distributed algorithm of the triad-ATRG is summarized in Table~\ref{tab:triad_atrg_algorithm}. 
In constructing the triad representation using the canonical form, we distribute the tensors $A$, $X$, $\sigma$ to GPU $1$ and $\sigma$, $Y$, $D$ to GPU $2$ separably. Since the structure is no longer canonical after constructing $\Gamma^{\text{Triad}}$,
all fundamental tensors as defined in Eqs.~\eqref{eq:all_fundamental_tensors_E}-~\eqref{eq:all_fundamental_tensors_L}
must be kept in order to derive squeezers.
To perform the SVD of $\Gamma^{\text{Triad}}$, we then compute the Gram matrices $O$ and $P$ for the upper block $(E, F, I, J)$ and the lower block $(G, H, K, L)$, respectively. In the squeezer step, we distribute the tensors $E$, $F$, $I$, $J$, and $P$ to GPU 1, and $G$, $H$, $K$, $L$ and $O$ to GPU 2.\footnote{In practice, each direction can be parallelized independently, allowing the use of up to six GPUs. However, since this part is not a bottleneck, such an implementation is not adopted in this work.}
In the coarse-graining process—which is the most computationally intensive step—we distribute the tensors $E$, $F$, $I$, $J$, and $G$, $H$, $K$, $L$ appearing in Eqs.~\eqref{makeMG} and \eqref{makeMH} across the available GPUs by partitioning the first index of tensors $E$ and $H$. 
It is straightforward to extend this algorithm to the ILO triad-ATRG and the ATRG methods by distributing the local tensors across the available GPUs. This parallelization technique was originally implemented in the HOTRG and ATRG methods~\cite{akiyamaPhaseTransitionFourdimensional2019, yamashitaParallelComputingMethod2022} for distributed CPU nodes, and we have successfully extended it to the GPU environment. 

In this study, two NVIDIA V100 GPUs were used for computations with small bond dimensions and for measuring computation time, whereas computations for large bond dimensions were carried out on eight NVIDIA A100 GPUs.
\begin{table}[H]
    \caption{Parallelized triad-ATRG algorithm on $n$ GPUs. Each box indicates the type of tensors distributed to each GPU. The notation $E(1{:}D/n,:,:,:)$ means that the first index of the tensor $E$ is restricted to the range from $1$ to $D/n$.}
    \label{tab:triad_atrg_algorithm}
    \centering
    \resizebox{\textwidth}{!}{
    \begin{tabular}{|c|c|c|c|c|c|}
        \hline
        GPU & Bond swapping & Make Triad rep. & Squeezer step & Making $\Phi$ & Making $\Psi$ \\
        \hhline{|=|=|=|=|=|=|}
        \hline
        GPU $1$ & $C, D$ & $A, X, \sigma$ & $E, F, I, J, P$ & $E(1:D/n,:,:,:), F, I, J, \{M\}$ & $H(1:D/n,:,:,:), G, K, L, \{N\}$ \\
        \hline
        GPU $2$ & None & $\sigma, Y, D$ & $G, H, K, L, O$ & $E(D/n+1:2D/n,:,:,:), F, I, J, \{M\}$ & $H(D/n+1:2D/n,:,:,:), G, K, L, \{N\}$ \\
        \hline
        GPU $3$ & None & None & None & $E(2D/n+1:3D/n,:,:,:), F, I, J, \{M\}$ & $H(2D/n+1:3D/n,:,:,:), G, K, L, \{N\}$ \\
        \hline
        $\vdots$ & $\vdots$ & $\vdots$ & $\vdots$ & $\vdots$ & $\vdots$ \\
        \hline
        GPU $n$ & None & None & None & $E((n-1)D/n+1:D,:,:,:), F, I, J, \{M\}$ & $H((n-1)D/n+1:D,:,:,:), G, K, L, \{N\}$ \\
        \hline
    \end{tabular}
    }
\end{table}
%
%
%

\section{Numerical results}
\label{sec:numerical_results}
In this section, the numerical benchmarking results of the triad-ATRG will be presented.
To this end, a comparison is made between the performance of the triad-ATRG and the ATRG using the four-dimensional Ising model. The Hamiltonian is given by
$H=-\sum_{\langle i,j\rangle}\sigma_i\sigma_j$ ($\sigma_i=\pm$)
where $\langle i,j\rangle$ denotes all possible pairs of nearest neighbor lattice sites.

The partition function of the four-dimensional Ising model
can be expressed as the tensor network form in Eq.~\eqref{eq:Z}.
It is well known that the tensor $T$ can be expressed by the canonical polyadic (CP) decomposition form with a matrix $W$ as
%
%
\begin{align}    &T_{i_1(n)i_2(n)i_3(n)i_4(n)j_1(n)j_2(n)j_3(n)j_4(n)}\cr
&=\sum_{a=1}^2 W_{ai_1(n)}W_{ai_2(n)}W_{ai_3(n)}W_{ai_4(n)}W_{aj_1(n)}W_{aj_2(n)}W_{aj_3(n)}W_{aj_4(n)}.
\end{align}
For the initial tensor $T$, the matrix $W$ is directly derived from the partition function $Z={\rm Tr}\left(e^{-\beta H}\right)$ with the Hamiltonian of the Ising model $H$ as 
%
%
\begin{equation}
    W=\begin{pmatrix}
        \sqrt{\cosh{\mathit{\beta}}} & \sqrt{\sinh{\mathit{\beta}}} \\
        \sqrt{\cosh{\mathit{\beta}}} & -\sqrt{\sinh{\mathit{\beta}}}\\
\end{pmatrix},
\end{equation}
where $\beta$ denotes the inverse temperature.

Previous studies of the four-dimensional Ising model using the tensor network method were conducted by the HOTRG and ATRG methods~\cite{Akiyama:2019chk,akiyamaPhaseTransitionFourdimensional2019}.
In this study, the triad-ATRG method is applied to the four-dimensional Ising model to demonstrate its performance.

\subsection{Demonstration of the computational time}
First of all, we examine the computational scaling of the triad-ATRG and ATRG methods with respect to the bond dimension $D$.
For this comparison, 
the free energy of the four-dimensional Ising model, $F=-\frac{1}{\beta}\log Z$, 
are evaluated in a volume $V=1024^4$, which is large enough to be considered the thermodynamic limit. For the triad-ATRG method, the overspanning parameter is chosen to be $r=7$.

\begin{figure}[tbp]
    \centering
    \includegraphics[width=0.65\textwidth]{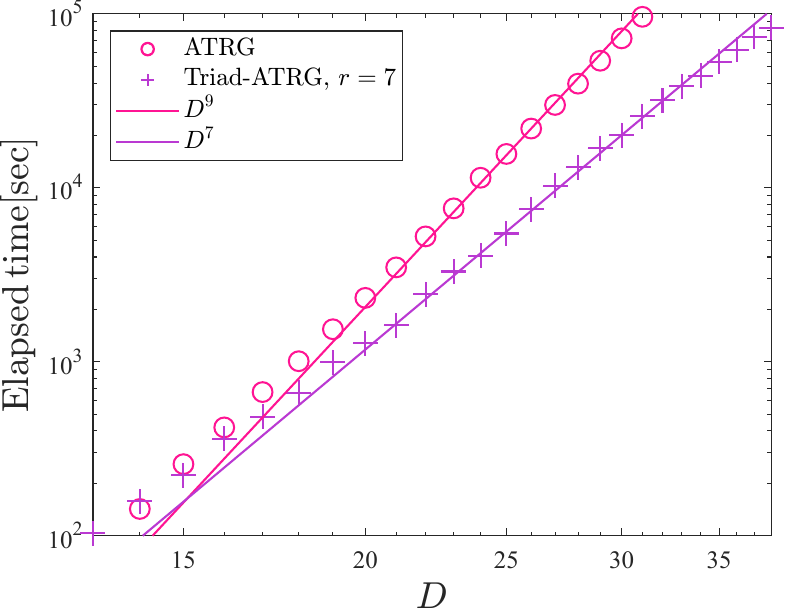}
    \caption{Computational time measured on a single CPU for the triad-ATRG and ATRG methods as {a function} of the bond dimension $D$.
    The magenta circles represent the results of the ATRG method, while the purple symbols represent the results of the triad-ATRG method with $r=7$. Both methods are executed with $q=20$. 
The magenta and purple solid lines correspond to the $O(D^9)$ and $O(D^7)$ scaling behaviors, respectively.
}
\label{fig::time}
\end{figure}

Figure~\ref{fig::time} shows the computational time of the triad-ATRG method ($r=7$) and the ATRG method, measured on a single CPU processor, as a function of the bond dimension $D$. In both methods, the number of iterations for the bond-swapping step is fixed at $q=20$.
The scaling behavior of the computational time clearly indicates that the triad-ATRG method exhibits superior performance compared to the ATRG methed. 

As discussed in the previous section, the computational cost of the triad-ATRG method is expected to scale as $O(D^7)$, whereas that of the ATRG method is expected to scale as $O(D^9)$. 
For the sake of reference,
the magenta and purple lines illustrate the expected scaling behavior, which the respective data points appear to follow approximately at larger bond dimensions.
This observation confirms that the triad-ATRG method has succeeded in significantly improving the computational efficiency compared to the ATRG method.

\begin{figure}[tbp]
    \begin{minipage}{0.48\textwidth}
        \centering
        \includegraphics[width=\linewidth]{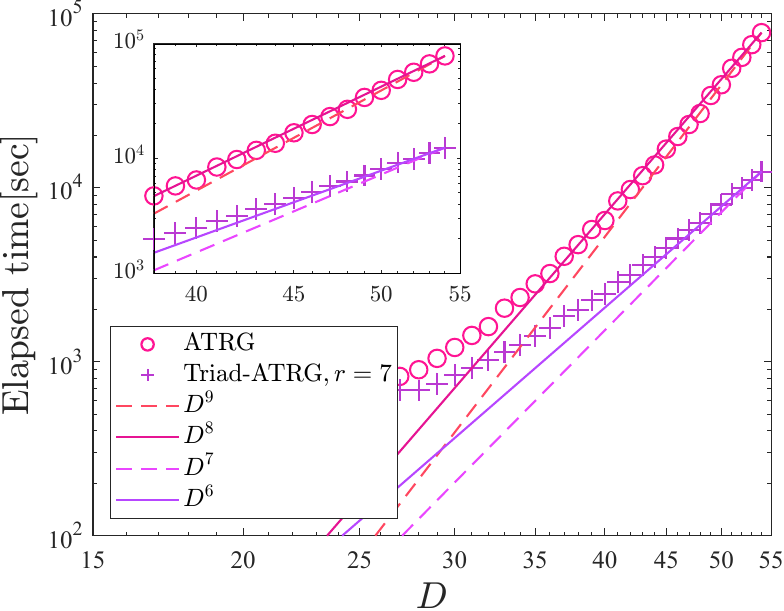}
        \refstepcounter{subfigure}\thesubfigure
        \label{fig:eff_a}
    \end{minipage}
    \hfill
    \begin{minipage}{0.48\textwidth}
        \centering
        \includegraphics[width=\linewidth]{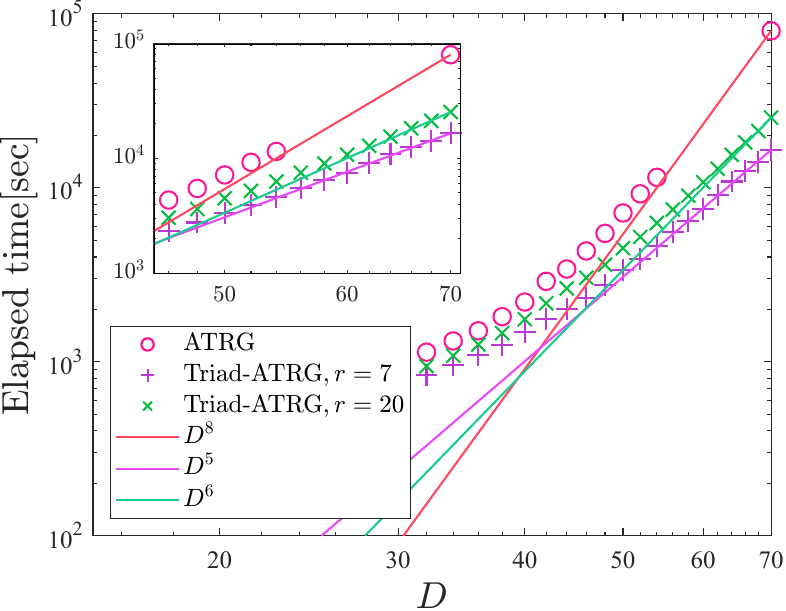}
        \refstepcounter{subfigure}\thesubfigure
        \label{fig:eff_b} 
    \end{minipage}
    \caption{Computational time for the ATRG and triad-ATRG methods as a function of the bond dimension~$D$.
    (a) Results are obtained using two NVIDIA V100 GPUs. The ATRG result is shown by magenta circles, while the triad-ATRG method with $r=7$ is shown by purple symbols. Solid lines indicate $O(D^9)$ (magenta) and $O(D^7)$ (purple) scalings, while dashed lines show $O(D^8)$ and $O(D^6)$ scalings for comparison. The inset magnifies the region $D \geq 38$ to highlight the asymptotic scaling.
    (b) Results are obtained using eight NVIDIA A100 GPUs. The ATRG result is again shown by magenta circles, the triad-ATRG result with $r=7$ and $r=20$ are indicated by purple and green symbols, respectively. Solid lines indicate $O(D^8)$ (magenta), $O(D^5)$ (purple), and $O(D^6)$ (green) scalings. The inset magnifies the region of $D \geq 48$.}
    \label{fig::timeGPU}
\end{figure}

Next, we examine the performance of these two methods when implemented on GPUs.
{Figure}~8.\ref{fig:eff_a} shows the computational times of the triad-ATRG ($r=7$) and ATRG methods measured using two NVIDIA V100 GPUs. The number of iterations for the bond-swapping step is set to be $q=O(D)$ for both methods.

It is clearly observed that both methods exhibit superior scaling in comparison to their respective theoretical expectations. The observed enhancement of efficiency is likely attributable to the high parallelization efficiency of the GPU implementation.
A similar behavior has also been reported for the HOTRG method when applied to a two-dimensional lattice~\cite{jhaGPUAccelerationTensorRenormalization2024}.  
The ATRG method exhibits approximately $O(D^{8\text{--}9})$ scaling, while the triad-ATRG method demonstrates $O(D^6)$ or even lower scaling, suggesting a good scalability of the triad-ATRG method in GPU-parallelized environments. 
In Fig.8~\ref{fig:eff_b},
we also present the results obtained in an environment with eight NVIDIA A100 GPUs. 
A significant difference in computation time is observed
between the ATRG and the triad-ATRG methods for the region $D \geq 50$ with $r=7$ and $r=20$. At $D=70$ with $r=20$, the triad-ATRG method has been demonstrated to be more than three times faster than the ATRG method. As observed in the case of the two-GPU system, the observed scaling behavior remains lower than the theoretical estimation. 
This behavior can be attributed to the fact that $D$ is not sufficiently large to reach the true asymptotic regime, thereby ensuring that lower-order contributions and overheads such as kernel launches remain non-negligible. Nevertheless, this does not affect the region of practical relevance for bond dimensions, since it is difficult to achieve very large bond dimensions in practice due to the rapidly increasing GPU memory requirements for the fundamental tensors $A,B,C$ and $D$, as well as the corresponding computational cost. Further investigations with $D \geq 70$ using a highly parallelized GPU implementation are required to clarify this point. We will address this in future work.

\subsection{Accuracy of the free energy density}

\begin{figure}[tbp]
    \centering
    \includegraphics[width=0.65\linewidth]{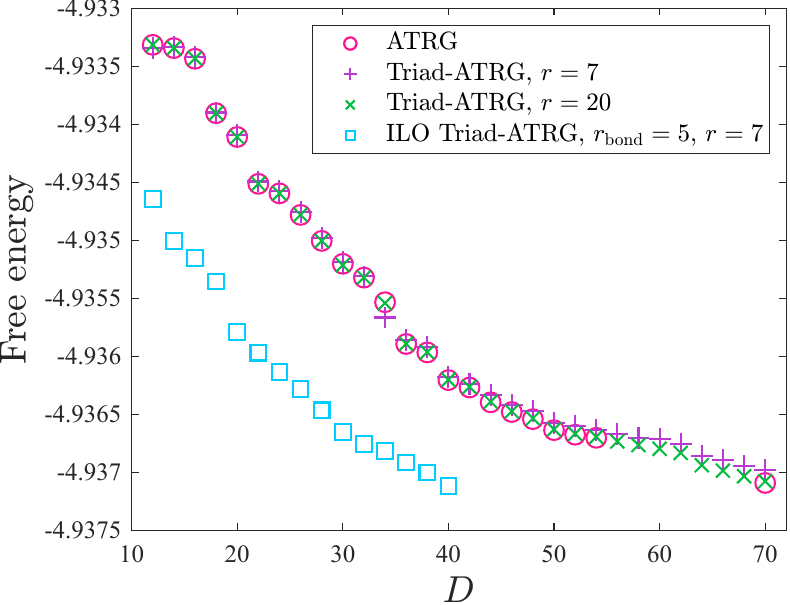}
    \caption{Free energy of the four-dimensional Ising model calculated by the ATRG, triad-ATRG, and ILO triad-ATRG methods. The ATRG results are denoted by magenta circles, while the triad-ATRG results with $r=7$ and $r=20$ are denoted by purple and green symbols, respectively. In addition to the two methods, the results of the ILO triad-ATRG method are also denoted by blue squares.
    }
    \label{fig::free_energy}
\end{figure}

Next let us verify the accuracy of the triad-ATRG method.
Figure~\ref{fig::free_energy} shows the free energy density $F/V$ of the four-dimensional Ising model near the critical point $T_c=6.65035$ reported in Refs.~\cite{Akiyama:2019chk,akiyamaPhaseTransitionFourdimensional2019}, and in the thermodynamic limit calculated by both the ATRG and triad-ATRG methods.  
The bond dimension $D$ is varied from $10$ to $54$ and $D=70$ for the ATRG, and from $10$ to $70$ for the triad-ATRG with overspanning parameter $r=7$ and $r=20$. 

Both the triad-ATRG and ATRG methods exhibit similar convergence behavior across the whole range of $D$. The results of the triad-ATRG show excellent agreement with those of the ATRG. 
For example, at $D=54$, the deviation from the ATRG result is only $0.0013\%$ for $r=7$, and $0.00015\%$ for $r=20$. Therefore, the triad-ATRG method can reproduce the ATRG results with sufficient accuracy.
From these results, it can be concluded that the triad-ATRG provides comparable accuracy to the ATRG in the evaluation of the free energy.
In addition, the triad-ATRG method can be more easily extended to larger bond dimensions, thanks to the reduction in computational cost.

Furthermore, the accuracy of the free energies calculated by the triad-ATRG method is significantly enhanced by the implementation of internal-line overspanning. 
As illustrated in Fig.~\ref{fig::free_energy}, the results obtained for the ILO triad-ATRG method with $r=7$ and $r_{\text{bond}}=5$ are overlaid.
It has been demonstrated that ILO yields a considerably lower free energy at the same bond dimension in comparison to the both the triad-ATRG and ATRG methods that do not employ overspanning in the bond-swapping step.

It should be emphasized that the free energy obtained by the ILO triad-ATRG even at $D=40$ is surprisingly lower than that of the triad-ATRG at $D=70$.
Based on this observation, overspanning in the bond-swapping step, where the approximation in the ATRG is most severe, leads to a significant improvement in accuracy.
Although the calculations for the ILO triad-ATRG method in this study have been performed on a single GPU, the current implementation of ILO makes it difficult to parallelize the procedure over multiple nodes or multiple GPUs due to the iterative use of QR decomposition and matrix multiplication in RSVD during the bond-swapping step.
Even when setting $r=1$, the computational cost of the bond-swapping step is still $O(r_{\text{bond}}^4 D^7)$, provided that $q = O(D)$. 
Therefore, we will need to develop a parallel implementation of the bond-swapping procedure on multiple GPUs in future work.

\subsection{Computational efficiency}

\begin{figure}[tbp]
    \centering
    
    \begin{minipage}{0.48\textwidth}
        \centering
        \includegraphics[width=\linewidth]{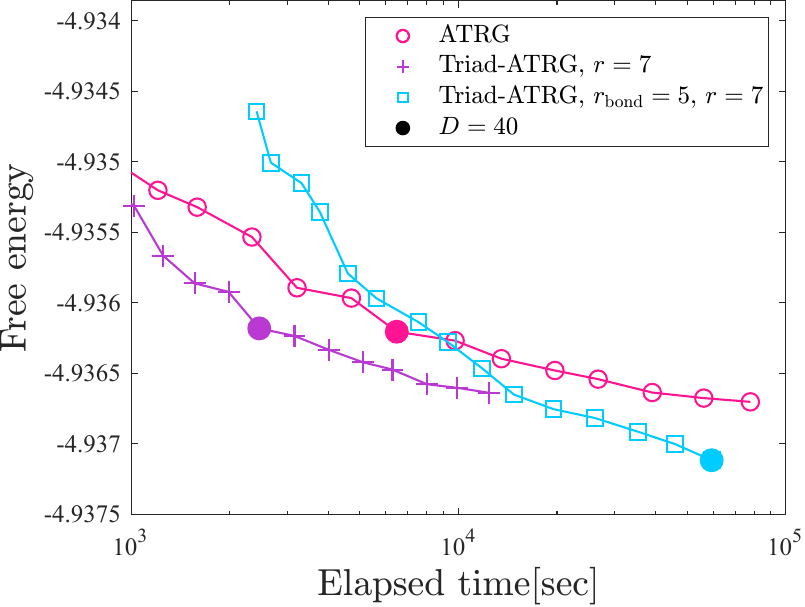}
        \refstepcounter{subfigure}\thesubfigure
        \label{fig:9_a}
    \end{minipage}
    \hfill
    \begin{minipage}{0.48\textwidth}
        \centering
        \includegraphics[width=\linewidth]{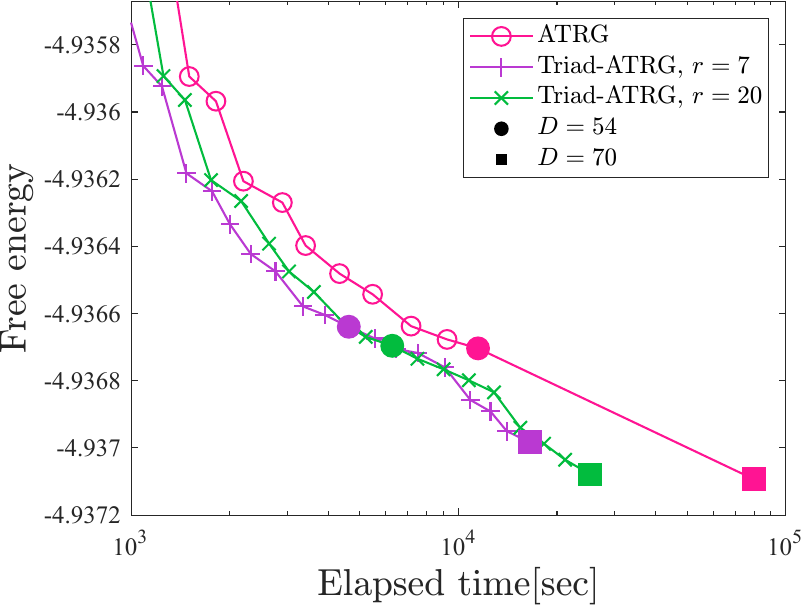}
        \refstepcounter{subfigure}\thesubfigure
        \label{fig:9_b}
    \end{minipage}
    \caption{
    Free energy as a function of the elapsed time measured on two NVIDIA V100 GPUs (left panel)
    and eight NVIDIA A100 GPUs (right panel).
    (a) Results are obtained using two NVIDIA V100 GPUs. 
    The pink, purple, and blue symbols represent results of the ATRG, triad-ATRG ($r=7$), and ILO triad-ATRG methods ($r_{\text{bond}}=5,\, r=7$), respectively. The filled circles indicate the data point at $D=40$. 
    (b) Results are obtained using eight NVIDIA A100 GPUs.
    The pink, purple, and green symbols represent results of the ATRG, triad-ATRG ($r=7$), and triad-ATRG ($r=20$) methods, respectively. The filled circles indicate the data point at $D=54$, and the filled squares correspond to $D=70$.}
    \label{fig::efficiency}
\end{figure}

In the previous subsections, we have shown that the triad-ATRG method achieves 
a level of accuracy comparable to that of the ATRG method for free energy, while demonstrating lower computational time and better scaling. Figure~\ref{fig::efficiency} shows the free energy as a function of elapsed time measured on the two NVIDIA V100 system (left panel) and the eight NVIDIA A100 system (right panel). It is clearly observed that the triad-ATRG method produce a lower free energy within a fixed computation time.
Regardless of which system is used, the triad-ATRG method becomes 
more efficient than the ATRG method in regions with longer elapsed times.

For comparison, we also provide the results of the ILO triad-ATRG method with $r_\text{bond}=5$ and $r=7$ when using the two-GPU system.

It is clear that for regions of elapsed time greater than $10^4$ sec, the ILO triad-ATRG method yields a significantly smaller free energy than the ATRG method.
For this reason, we conclude that both the triad-ATRG and ILO triad-ATRG methods are more efficient than
the ATRG method for calculating free energy in GPU environments.

\subsection{Comparison of other physical quantities}
Finally, we compare the physical quantities calculated by the ATRG and triad-ATRG methods, such as the transition temperature $T_c$ and internal energy $U$ in this subsection.

We use the following dimensionless parameter proposed in Ref.~\cite{guTensorEntanglementFilteringRenormalizationApproach2009} to determine the transition point $T_c$,
%
%
\begin{equation}
X^{(m)}=\frac{(\mathrm{Tr}A^{(m)})^2}{\mathrm{Tr}(A^{(m)})^2},\text{ with }A^{(m)}_{kl}=\sum T_{i_1 i_2 i_3 k i_1 i_2 i_3 l}^{(m)}
\end{equation}
where $T^{(m)}$ denotes the renormalized tensor at the $m$-th iteration. It is known that the quantity $X$ can effectively count the degeneracy of the ground state.
Since $X$ is expected to be 2 in the ordered phase and 1 in the disordered phase, the point at which the quantity $X$ abruptly drops from $2$ to $1$ can be identified as the phase transition point.

\begin{figure}[tbp]
    \centering
    \includegraphics[width=0.65\linewidth]{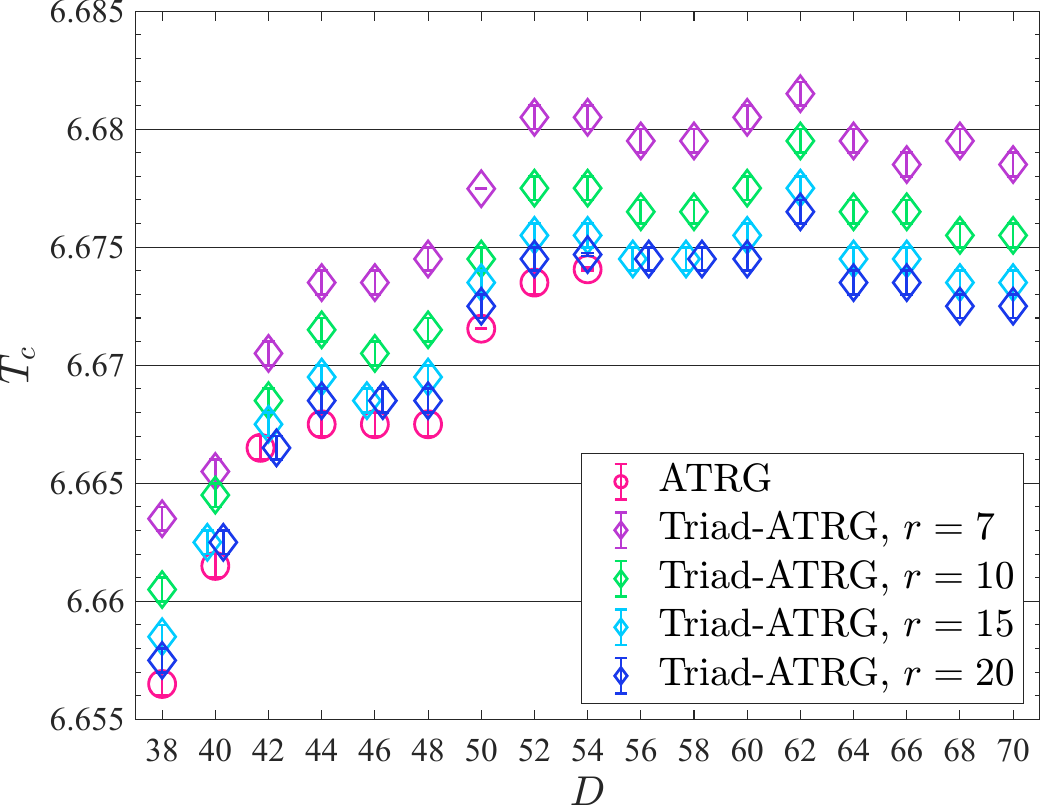}
    \caption{Transition temperature $T_c$ of the four-dimensional Ising model as a function of $D$, calculated by the ATRG and triad-ATRG methods with $r=7,\,10,\,15,\,20$. The ATRG results are shown in magenta circle, while the triad-ATRG results are {represented by purple ($r=7$), green ($r=10$), cyan($r=15$), and blue ($r=20$) symbols,} respectively.}
    \label{fig::Tc}
\end{figure}

\begin{figure}[tbp]    
    \centering
    \includegraphics[width=0.65\linewidth]{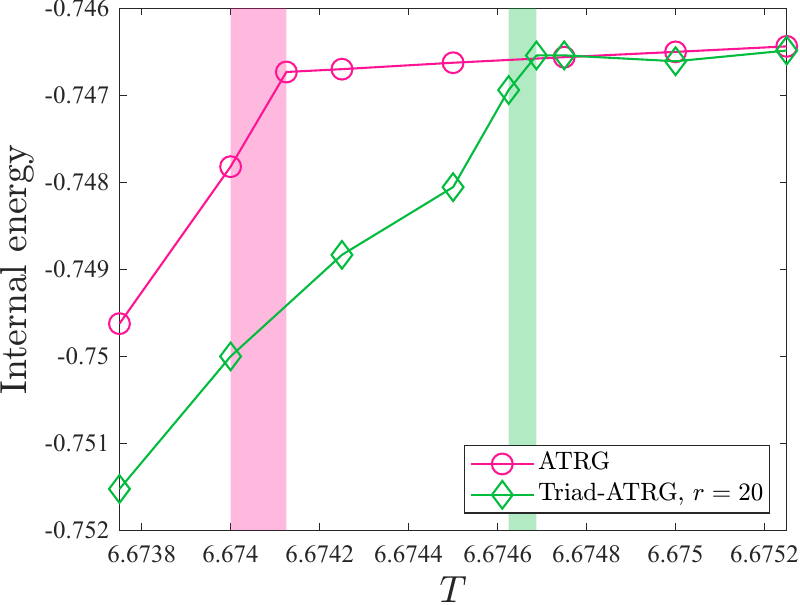}
    \caption{Comparison of internal energy between the triad-ATRG ($r=20$) and ATRG methods for bond dimension $D=54$. The horizontal axis represents temperature. The purple and green horizontal bands represent the phase transition points obtained from
    the ATRG and triad-ATRG methods, respectively.}
    \label{internalenergy}
\end{figure}

To investigate how the phase transition point depends on the bond dimension $D$, we plot the transition temperature $T_c$ as a function of $D$ for the ATRG, and triad-ATRG with $r=7$, 10, 15, and 20 in Fig.~\ref{fig::Tc}. Error bars indicate the resolution of the critical temperature search used in this calculation.
It has been demonstrated that the triad-ATRG and ATRG methods produce results that are analogous, with minor deviations depending on the choice of the overspanning parameter $r$.
By increasing $r$, transition points are shifted to lower values, and asymptotically approach the ATRG results.
These results show that, at $D=54$, the difference between the ATRG and triad-ATRG methods is approximately $0.1\%$ for $r=7$, about $0.05\%$ for $r=10$, and reduces to around $0.01\%$ for $r=20$. 
Of particular significance is the observation that the triad-ATRG method yields results up to $D=70$, with clearly better convergence with respect to $D$ that results up to $D=54$. 
It should be noted that the estimated value of $T_c$ obtained in this study is relatively lower than the previous Monte Carlo result recently reported in Ref.~\cite{Lundow:2022ltb} and higher than the previous result provided by the HOTRG and ATRG method~\cite{Akiyama:2019chk,akiyamaPhaseTransitionFourdimensional2019}.
The physical quantities previously examined, including the free energy and the critical temperature, can be calculated from the pure tensor network, as defined in Eq.~\eqref{eq:Z}. On the other hand, the internal energy of the four-dimensional Ising model can be computed by contracting the tensor network with two impurity tensors~\cite{akiyamaPhaseTransitionFourdimensional2019}.  
We then compare the accuracy of the internal energy obtained by the two methods using the tensor network including impurity tensors.
Figure~\ref{internalenergy} shows the internal energy computed using both the ATRG and triad-ATRG methods with bond dimension $D=54$. 
The overspanning parameter for the triad-ATRG method is set to be $r=20$.

The purple and green horizontal bands represent the phase transition points obtained from $X$ for both methods.
Although the two results show slightly different behavior at low temperatures due to a shift in the phase transition point, the numerical discrepancy is not significant.
The difference is only $0.002\%$ at high temperature ($T=6.67475$), and $0.2\%$ at low temperature ($T=6.674$).
These results suggest that the observed differences are sufficiently small. 
Therefore,
we conclude that the triad-ATRG method is as accurate as the ATRG method, even when the tensor network includes impurity tensors.

\section{Conclusion}
\label{sec:conclusion}
We have proposed a novel method called the triad-ATRG that significantly reduces the computational cost of the TRG approach in four dimensions. 
This reduction is achieved by combining features of both the ATRG~\cite{adachiAnisotropicTensorRenormalization2020} and MDTRG~\cite{nakayamaApplicationProjectiveTruncation2024,nakayamaRandomizedHigherorderTensor2023} methods.
The triad-ATRG achieves a lower computational scaling of $O(D^7)$,
for the four-dimensional case, compared to the original ATRG method, which scales as $O(D^9)$.
We have also confirmed that the triad-ATRG method maintains a comparable level of accuracy for the free energy and other physical quantities in comparison to the original ATRG method.
Furthermore, the accuracy can be further improved by implementing internal-line overspanning in the bond-swapping step, which we refer to as the ILO triad-ATRG. 
The implementation of the parallelized versions of the ATRG and triad-ATRG methods on multiple GPUs has also demonstrated very good scalability with respect to the bond dimension.
These results indicate that the triad-ATRG method is a promising method, offering lower computational costs and higher accuracy.
Therefore, our findings pave the way for applying the TRG method to various four-dimensional lattice models.

\begin{acknowledgments}
We {would like to} thank S. Akiyama and K. Nakayama for their helpful comments, fruitful discussions, and for {kindly} providing us with numerical data.
Y.S. is supported by Graduate Program on Physics for the Universe (GP-PU), Tohoku University and JSPS KAKENHI Grant 25KJ0537.
Numerical calculations were carried out on {Yukawa-21 at YITP in Kyoto University and SQUID at D3 Center in the University of Osaka. The tensor contractions on both CPU
and GPU were performed using \texttt{TensorOperations.jl}~\cite{tensoroperations}.}
\end{acknowledgments}
\appendix
\section{Implementation of the bond-swapping procedure}\label{append:A}
We review the algorithm of the bond-swapping procedure based on the randomized singular value decomposition (RSVD).
This technique was first introduced in Ref.~\cite{10.1093/ptep/ptz133} and later refined in Ref.~\cite{oai:tsukuba.repo.nii.ac.jp:02005504}.

The following discussion will examine the overspanning of internal lines by $r_\text{bond}$ times $D$ in the bond dimension, during the bond-swapping step of the ATRG method, with a focus on its application to a four-dimensional system.
When $r_\text{bond}=1$ is set, it returns to the conventional ATRG bond-swapping process.

As explained in Sec.~\ref{sec:triad_representation}, the bond-swapping step is the procedure that performs the truncated SVD of the sub-network $BC$,
%
%
\begin{align}\label{eq:bond-swapping}
\sum_{{i_1(n)}}^{r_\text{bond}D} B_{i_1(n)j_2(n+\hat{1})j_3(n+\hat{1})j_4(n+\hat{1})\alpha}C_{i_1(n)i_2(n)i_3(n) i_4(n)\vphantom{i_1(n+\hat{1})}\beta}\nonumber\\
    \approx \sum_{\gamma}^{r_\text{bond}D} X_{\alpha i_2(n)i_3(n)i_4(n)\vphantom{i_1(n+\hat{1})}\gamma}
\sigma_{\tsp\gamma\gamma}Y_{\beta j_2(n+\hat{1})j_3(n+\hat{1}) j_4(n+\hat{1})\gamma}.
\end{align}
Note that subscripts $\alpha$ and $\beta$ of $B$ and $C$ are already overspanned from $D$ to $r_\text{bond}D$.
It should be emphasized that this decomposition corresponds to yields an optimal low-rank approximation {of} the unit-cell tensor $\Gamma$ in terms of the Frobenius norm, since the tensors $A$ and $B$ defined in Eq.~\eqref{eq:ABCD} are isometric.
However, a naive implementation of this SVD requires $O(D^{2d+1})$ computational cost and $O(D^{2d})$ memory cost, {both of which are} extremely high. 

\begin{figure}[t]    
    \begin{center}
    \includegraphics[scale=1.6]{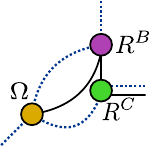}
    \caption{{Schematic view of the} tensor network {given} in Eq.~\eqref{eq:bond1}. Blue dotted lines are overspanned {from $D$} to $r_\text{bond}D$. This contraction {has an} $O(r_\text{bond}^4D^6)$ computational cost.}
    \label{fig:overbond}
    \end{center}
\end{figure}

To mitigate this cost, the RSVD-based method introduced in Ref.~\cite{10.1093/ptep/ptz133,oai:tsukuba.repo.nii.ac.jp:02005504} begins by applying QR (or singular value) decompositions to the tensors $B$ and $C$, 
%
%
\begin{align}
B_{i_1(n)j_2(n+\hat{1})j_3(n+\hat{1})j_4(n+\hat{1})\alpha}&=\sum_{p}^{r_\text{bond}D^2}Q_{j_2(n+\hat{1})j_3(n+\hat{1}) j_4(n+\hat{1})p}^B R_{i_1(n)\alpha p}^B\\
    C_{i_1(n)i_2(n)i_3(n)i_4(n)\vphantom{i_1(n+\hat{1})}\beta}&=\sum_{q}^{r_\text{bond}D^2}Q_{i_2(n)i_3(n)i_4(n)\vphantom{i_1(n+\hat{1})}q}^C R_{i_1(n)\beta q\vphantom{i_1(n+\hat{1})}}^C,
\end{align}
which can reduce the dimensionality of the tensors involved in the SVD of Eq.~\eqref{eq:bond-swapping}. In the RSVD procedure, we iteratively perform the following QR-based orthonormalization~\cite{doi:10.1137/090771806}:
%
%
\begin{align}
    &\Omega^{(0)}_{\beta p m} := \Omega_{\beta p m}, \nonumber \\
    &\text{for $k$ in $1:q$} \nonumber \\
    &\quad Y^{(k)}_{\alpha q m'} = \sum_{i_1(n),\,\beta,\,p}^{D,r_\text{bond}D,r_\text{bond}D^2}
    R^C_{i_1(n)\beta q} \left( R^B_{i_1(n)\alpha p} \, \Omega^{(k-1)}_{\beta p m} \right)\label{eq:bond1}\\
    &\quad \text{QR decomposition: } Y^{(k)} = Q^{(k)} R^{(k)} \nonumber \\
    &\quad \tilde{Y}^{(k)}_{\beta p m} = \sum_{i_1(n),\,\alpha,\,q}^{D,r_\text{bond}D,r_\text{bond}D^2}
    R^{*B}_{i_1(n)\alpha p} \left( R^{*C}_{i_1(n)\beta q} \, Q^{(k)}_{\alpha q m'} \right)\label{eq:bond2} \\
    &\quad \text{QR decomposition: } \tilde{Y}^{(k)} = Q'^{(k)} R'^{(k)}\nonumber \\
    &\quad \text{Update:} \quad \Omega^{(k)} := Q'^{(k)}.
\end{align}
In the above algorithm, $\Omega_{\beta p m}$ is a random tensor of {size} $(r_\text{bond}D) \times (r_\text{bond}D^2) \times (\eta D)$~\footnote{In practice, the index $m$ of the random tensor $\Omega$ is oversampled to size $\eta=r_\text{bond} + 1$ for the ILO triad-ATRG in this work, following the standard convention for the oversampling parameter in RSVD.}. 
The role of $Q^{(q)}$ is to capture the dominant subspace of $\Gamma$.  
By projecting onto the subspace spanned by $Q^{(q)}$, \emph{i.e.}, applying the projection operator $Q^{(q)} Q^{(q)\dagger}$, the truncated SVD is approximately realized as follows:
%
%
\begin{align}
\sum_{i_1(n)}^{D} R_{i_1(n)\alpha p}^B R_{i_1(n)\beta q\vphantom{i_1(n+\hat{1})}}^C &\approx \sum_{i_1(n),\alpha',q',m'}^{D,r_\text{bond}D,r_\text{bond}D^2,r_\text{bond}D} Q_{\tsp\alpha q m'}\left\{\left(Q_{\tsp\alpha'q' m'}^*R_{i_1(n)\beta q'\vphantom{i_1(n+\hat{1})}}^C\right)R_{i_1(n)\alpha' p}^B\right\}\\
    &\approx \sum_{m'\gamma}^{r_\text{bond}D,r_\text{bond}D} Q_{\tsp\alpha q m'}\tilde{U}_{m'\gamma}\sigma_{\gamma\gamma}V_{\beta p \gamma}^{{*\text{bond}}}\\
    &=\sum_{\gamma}^{r_\text{bond}D} U_{\alpha q \gamma}^{{\text{bond}}}\sigma_{\gamma\gamma}V_{\beta p \gamma}^{{*\text{bond}}}.
\end{align}
In the second line of the above equation, $\tilde{U}_{m'\gamma}$, $\sigma_{\gamma\gamma}$, and $V_{\beta p \gamma}^{*\text{bond}}$ are, the left isometry, the singular values, and the right isometry, {respectively.
They are} obtained from the SVD of the projected tensor $Q^{(q)\dagger} R^C R^B$.  
In the third line, we redefine $U^{\text{bond}} = Q^{(q)} \tilde{U}$ to absorb the isometric basis $Q^{(q)}$ into the left unitary.
By using these tensors, we can define 
%
%
\begin{align}\label{XY}
    &X_{\alpha i_2(n)i_3(n) i_4(n)\vphantom{i_1(n+\hat{1})}\gamma}=\sum_{q}^{r_\text{bond}D^2}U_{\alpha q \gamma}^{{\text{bond}}}Q_{i_2(n)i_3(n)i_4(n)\vphantom{i_1(n+\hat{1})}q}^C\\
    &Y_{\beta j_2(n+\hat{1})j_3(n+\hat{1})j_4(n+\hat{1})\gamma}=\sum_{p}^{r_\text{bond}D^2}V_{\beta p\gamma}^{{*\text{bond}}} Q_{j_2(n+\hat{1})j_3(n+\hat{1})j_4(n+\hat{1})p}^B.\label{XY2}
\end{align}
The bottleneck of the bond-swapping step lies in the contractions in Eq.~\eqref{eq:bond1} and Eq.~\eqref{eq:bond2}, {as shown} in Fig.~\ref{fig:overbond}.  
{These contractions} must be performed $q$ times, {each requiring
an $O(r_\text{bond}^4 D^6)$ computational cost.}  
Thus, if we set $q = O(D)$, the overall cost becomes $O(r_\text{bond}^4 D^7)$, {comparable to the cost} of the coarse-graining step.
In practical computations of the ILO triad-ATRG, {the tensors $X$ and $Y$ are not explicitly constructed.}
Instead, we {work directly} with $U^{\text{bond}}$, $\sigma$, and $V^{*\text{bond}}$ to {create} isometries of the triad representation. {Therefore,} $O(r_\text{bond}^2D^5)$ memory cost is avoided.
\bibliography{TRG}
\end{document}